\documentstyle[12pt]{article}
\begin{document}

\title{\bf Entropy of Extremal Black Holes in Two Dimension }
\author{{J. Sadeghi $^{a}$\thanks{Email: pouriya@ipm.ir}\hspace{1mm}
M. R. Setare $^{b}$ \thanks{Email: rezakord@ipm.ir} and B.
Pourhassan$^{a}$}\\
{ $^{a}$ Sciences Faculty, Department of Physics, Mazandaran
University,}\\{  P .O .Box 47415-416, Babolsar, Iran} \\ { $^b$
Department of Science, Payame Noor University, Bijar, Iran}}

 \maketitle

\begin{abstract}
In this paper we apply the entropy function formalism to the
two-dimensional black hole which  come from the compactification of
the heterotic string theory with the dilaton coupling function. We
find the Bekenstein-Hawking entropy  from the value of the entropy
function at its saddle point. Also we consider higher derivative
terms. After that we apply the entropy function formalism to the
Jackiw-Teitelboim (JT) model where we consider the effect of
string-loop to this model.
\end{abstract}
\newpage
\section{Introduction}
The entropy function method is appropriate way to determined the
entropy of two-dimensional black hole, because in two-dimensional
black hole the  horizon is a point, so horizon area simply vanishes
and  seems entropy must to be zero. However in two-dimensional
dilaton gravities [1, 2, 3, 4], it has been shown that the entropy
is proportional to the value of the dilaton field at the horizon.
Recently people find that the entropy function [5] can derived from
Wald's formula [6], for these aim one first rewrite the lagrangian
density in terms of value of fields near horizon, and then taking
the Legendre transform of the resulting function with respect to the
electric field [6,7]. The near horizon geometry of the extremal
black hole is determined by extremizing the entropy function and the
black hole entropy is given by the extremum value of the entropy
function. This general method is an easier way to calculate the
black hole entropy.\\
Entropy function analysis provides a good understanding of the
attractor
mechanism for spherically symmetric extremal black holes if:\\
1.We consider a theory of gravity coupled to abelian (p-form) gauge
fields and neutral scalar fields.\\
2.The Lagrangian density $f$ is gauge and general coordinate
invariant.\\
3.Define an extremal black hole to be one whose near horizon
geometry is $AdS_{2}\times S^{2}$ (in $D = 4$).\\
In this approach, the theory need not be supersymmetric and $f$
could contain higher derivative terms. For such black holes one can
define an 'entropy function' $F$ as follows:
\begin{equation}\label{m1}
F=2\pi[q_{i}\epsilon_{i}-f]
\end{equation}
where $q_{i}$ denote electric charges, and $\epsilon_{i}$ are near
horizon radial electric field. $F$ is a function of the $q_{i}$  and
various parameters labeling the $SO(2,1)\times SO(3)$ symmetric near
horizon background (e.g. sizes of $AdS_{2}$ and $S^{2}$,  vacuum
expectation value of scalars, radial electric fields, radial
magnetic fields ). Then for a black hole with given electric charges
$q$ and magnetic charges $p$, all other near horizon parameters are
obtained by extremizing $F$ with respect to these parameters. And
finally the entropy is given by the value of $F$ at its extremum.\\
 So, in section two we calculate the entropy
function for two-dimensional effective heterotic string theory. In
that case the function of the dilaton $\Phi$ appearing as a common
factor of the corresponding action is given by some series. By
considering this series we account the string tree level
contribution [9, 10, 11] and string-loop affect to the entropy
function and obtained the entropy of system. In section (3) we apply
the entropy function formalism to the Jackiw-Teitelboim (JT)
model[3].

\section{Two-dimensional effective heterotic string theory}
First we consider the two-dimensional gravity, which may come from
the compactification of the heterotic string
 theory with the dilaton coupling function $B(\Phi)$, $U(1)$ gauge field $F$ and the two-dimensional cosmological constant $\lambda$. The action is given by,
\begin{equation}\label{m1}
S=\frac{1}{2\pi}\int{d^{2}x\sqrt{-g}B(\Phi)\left[R+4(\nabla\Phi)^{2}+4\lambda^{2}-\frac{F^{2}}{4}\right]},
\end{equation}
where
\begin{equation}\label{m2}
B(\Phi)=e^{-2\Phi}+C_{0}+C_{1}e^{2\Phi}+C_{2}e^{4\Phi}+\cdots
\end{equation}
The first term on the right-hand side of Eq.(\ref{m2}), is the
string tree level contribution \cite{{c10},{c11}}. The higher order
terms represent the string-loop effects. Apart from the fact that
$B(\Phi)$ is a series in powers of $exp(2\Phi)$,  little is known
about the global behavior of the dilaton coupling function
$B(\Phi)$. Now we are going to apply this theory  to the
Callan-Giddings-Harvey-Strominger (CGHS) model and study the quantum
nature of the black hole.  So, The near horizon solutions of the
static charged black hole which has $SO(2, 1)$ symmetry can be
written as
\begin{eqnarray}\label{m3}
ds^{2}&=&v(-r^{2}dt^{2}+\frac{1}{r^{2}}dr^{2}),\nonumber\\
\phi&=&u,\nonumber\\
F_{rt}&=&\epsilon,
\end{eqnarray}
where $\phi=e^{-2\Phi}$.\\
Here  $u$, $v$ and $\epsilon$ are constants, where can  be
determined in terms of the charge $q$ and cosmological constant
$\lambda$. Note that the covariant derivatives of the Riemann
tensor, the scalar field and the gauge field strength all vanish in
this near horizon geometry. This plays an important role to
construct Sen's entropy function from Wald's formula.
 Therefore  by using equation (\ref{m3})  the Lagrangian density becomes,
\begin{equation}\label{m4}
f(u, v, \epsilon)=\frac{v}{2\pi}B(u)\left[-\frac{2}{v}+4\lambda^{2}+\frac{\epsilon^{2}}{2v^{2}}\right],
\end{equation}
where $B(u)=u+C_{0}+\frac{C_{1}}{u}+\cdots$. Here we kept just three terms from dilaton coupling function.\\
From equation (4) one can finds the electric charged as,
\begin{equation}\label{m5}
q=\frac{\partial f}{\partial \epsilon}=\frac{\epsilon B(u)}{2\pi v}.
\end{equation}
Now, the entropy function is defined as the Legendre transformation of
the Lagrangian density with respect to the gauge field $\epsilon$,
\begin{equation}\label{m6}
F(u, v, q)=2\pi[q\epsilon-f]=vB(u)\left[\frac{2}{v}-4\lambda^{2}+\frac{2\pi^{2}q^{2}}{B^{2}(u)}\right].
\end{equation}
The undetermined parameter $u$ and $v$ can be fixed by the equations
of motion, which becomes  the extremum equations as,
\begin{equation}\label{m7}
\frac{\partial F}{\partial v}(u_{e}, v_{e})=\left[-4\lambda^{2}B(u_{e})+\frac{2\pi^{2}q^{2}}{B(u_{e})}\right]=0,
\end{equation}
and
\begin{equation}\label{m8}
\frac{\partial F}{\partial u}(u_{e}, v_{e})=\left[2-4\lambda^{2}v_{e}-\frac{2\pi^{2}q^{2}v_{e}}{B^{2}(u_{e})}\right]=0.
\end{equation}
Equations (\ref{m7}) and (\ref{m8}) yields the solutions as
following,
\begin{eqnarray}\label{m9}
u_{e}&=&\frac{1}{2}\left((\frac{\pi q}{\sqrt{2}\lambda}-C_{0})+\sqrt{(\frac{\pi q}{\sqrt{2}\lambda}-C_{0})^{2}-4C_{1}}\right),\nonumber\\
v_{e}&=&\frac{1}{4\lambda^{2}}.
\end{eqnarray}
The entropy is given by the value of the entropy function at the extremum,
\begin{equation}\label{m10}
S_{BH}(q)=F(u_{e}, v_{e}, q)=2(u_{e}+C_{0}+\frac{C_{1}}{u_{e}}).
\end{equation}
Therefore the string-loop correction give us the entropy as equation (\ref{m10}).\\
Now let us consider the effect of higher derivative terms. Since in
two dimensions, Riemann and Ricci tensors can be expressed in terms
of Ricci scalar, it is sufficient to consider the higher derivative
terms of the form $R^{n}$. This means we must do replacement
$R\rightarrow \sum{a_{n}R^{n}}=R+a_{2}R^{2}+\cdots$. Due to the
higher derivative terms the corresponding action can be written by
\begin{equation}\label{m11}
S=\frac{1}{2\pi}\int d^{2}x{\sqrt{-g}B(\Phi)\left[R+\sum{a_{n}R^{n}}+ 4(\nabla\Phi)^{2}+4\lambda^{2}-\frac{F^{2}}{4}\right]},
\end{equation}
so the entropy function is as following,
\begin{equation}\label{m12}
F=vB(u)\left[\frac{2}{v}-4\lambda^{2}+\frac{2\pi^{2}q^{2}}{B^{2}(u)}-\sum {a_{n}(\frac{-2}{v})^{n}}\right].
\end{equation}
We note that the entropy (\ref{m10}) is modified as,
\begin{equation}\label{m13}
S_{mod}=B(u_{e})\left[2-\sum n a_{n}(-2)^{n}v^{1-n}_{e}\right].
\end{equation}
Also equations (\ref{m7}) and (\ref{m8}) are modified as
\begin{equation}\label{m14}
\frac{\partial F}{\partial v}(u_{e}, v_{e})=\left[-4\lambda^{2}B(u_{e})+\frac{2\pi^{2}q^{2}}{B(u_{e})}-B(u_{e})\sum (1-n) a_{n}(-2)^{n}v^{-n}_{e}\right]=0,
\end{equation}
and
\begin{equation}\label{m15}
\frac{\partial F}{\partial u}(u_{e}, v_{e})=\left[2-4\lambda^{2}v_{e}-\frac{2\pi^{2}q^{2}v_{e}}{B^{2}(u_{e})}-\sum a_{n}(-2)^{n}v^{1-n}_{e}\right]=0.
\end{equation}
Using these equations, the entropy (\ref{m13}) can be written in
terms of $q$,
\begin{equation}\label{m16}
S_{mod}=\frac{4\pi^{2}q^{2}v_{e}}{(u_{e}+C_{0}+\frac{C_{1}}{u_{e}})}.
\end{equation}
\section{ Jackiw-Teitelboim (JT) model }
Another interesting model in two-dimensional gravity is the
Jackiw-Teitelboim (JT) model \cite{c3}. We add the effect of
string-loop to this model. Moreover in order to study the extremal
charged black hole, one can include a gauge field in the model with
the following form of the action,
\begin{equation}\label{m17}
S=\frac{1}{2\pi}\int{d^{2}x\sqrt{-g}B(\Phi)\left[R+4(\nabla\Phi)^{2}+4\lambda^{2}-B^{2}(\Phi)\frac{F^{2}}{4}\right]}.
\end{equation}
The near horizon solutions are same equation (\ref{m3}). Therefore
one can study the black hole entropy in the similar way. As before
the entropy function $F$ is the Legendre transform of the Lagrangian
density $f$ in terms of the electric field.
\begin{equation}\label{m18}
F(u, v, q)=vB(u)\left[\frac{2}{v}-4\lambda^{2}+\frac{2\pi^{2}q^{2}}{B^{4}(u)}\right],
\end{equation}
where the electric charge is given by,
\begin{equation}\label{m19}
q=\frac{\epsilon B^{3}(u)}{2\pi v}.
\end{equation}
Extremizing the entropy function with respect to $v$ and $u$,
\begin{equation}\label{m20}
\frac{\partial F}{\partial v}(u_{e}, v_{e})=\left[-4\lambda^{2}B(u_{e})+\frac{2\pi^{2}q^{2}}{B^{3}(u_{e})}\right]=0,
\end{equation}
and
\begin{equation}\label{m21}
\frac{\partial F}{\partial u}(u_{e}, v_{e})=\left[2-4\lambda^{2}v_{e}-\frac{6\pi^{2}q^{2}v_{e}}{B^{4}(u_{e})}\right]=0,
\end{equation}
provides the extremizing solutions,
\begin{eqnarray}\label{m22}
u_{e}&=&\frac{1}{2}\left((\sqrt{\frac{\pi q}{\sqrt{2}\lambda}}-C_{0})+\sqrt{(\sqrt{\frac{\pi q}{\sqrt{2}\lambda}}-C_{0})^{2}-4C_{1}}\right),\nonumber\\
v_{e}&=&\frac{1}{8\lambda^{2}}.
\end{eqnarray}
By plugging these back into the entropy function, we obtain the black hole entropy as
\begin{equation}\label{m23}
S_{BH}(q)=2(u_{e}+C_{0}+\frac{C_{1}}{u_{e}})=\sqrt{\frac{\pi q}{\sqrt{2}\lambda}}.
\end{equation}
We can consider the higher derivative corrections in the similar
way. In that case one can find,
\begin{equation}\label{m24}
S_{mod}=\frac{4\pi^{2}q^{2}v_{e}}{(u_{e}+C_{0}+\frac{C_{1}}{u_{e}})^{4}}.
\end{equation}
\section{Conclusion}
The entropy formalism, describes the attractor equations and black
hole entropy in a general non-supersymmetric and higher derivative
gravity theory. In this formalism, the near horizon geometry is
determined by extremizing a single function $F$, the entropy
function. The entropy of the black hole is given by the value of $F$
at the extremum. In this paper we use the entropy function to obtain
entropy of two-dimensional charged black hole. We discussed about
two models in dilaton gravity then we considered the effect of
string loop corrections in both models . Finally we obtained
modified entropy under effect of higher derivative terms. In this
work we considered just three terms of dilaton coupling function.
One can include more than terms and find same entropy.


\begin{thebibliography}{11}
\bibitem{c1}
S. W. Hawking, \textit{Comm. Math Phys}. \textbf{43}, 199 (1975).
\bibitem{c2}
C. G. Callan, S. B. Giddings, J. A. Harvey, and A. Strominger,
\textit{Phys. Rev. D} \textbf{45}, 1005 (1992).
\bibitem{c3}
R. Jackiw, \textit{Nucl. Phys. B} \textbf{252}, 343 (1985).
\bibitem{c4}
C. Teitelboim,\textit{Phys. Lett. B} \textbf{126}, 41 (1983).
\bibitem{c5}
A. Sen,\textit{JHEP} \textbf{0509}, 038 (2005).
\bibitem{c6}
R. M. Wald, \textit{Phys. Rev. D} {\bf 48}, 3427, (1993).
\bibitem{c7}
V. Iyer and R. M. Wald, \textit{Phys. Rev. D} \textbf{50},
846 (1994)
\bibitem{c9}
E. S. Fradkin and A. A. Tseytlin, \textit{Phys. Lett. B} \textbf{158}, 316 (1985).
\bibitem{c10}
C. G. Callan, D. Friedan, E. J. Martinec and M. J. Perry,
\emph{Nucl. Phys. B} \textbf{262}, 593(1985).
\bibitem{c11}
C. G. Callan, I. R. Klebanov and M. J. Perry, \textit{Nucl. Phys. B}
\textbf{278}, 78 (1986).

\end{thebibliography}
\end{document}